\pdfoutput=1
\documentclass[aps,prb,floatfix,superscriptaddress,reprint]{revtex4-2}
\usepackage{eurosym}
\usepackage{amsbsy}
\usepackage{array}
\usepackage{pifont}
\usepackage{latexsym,epsfig,graphicx}
\usepackage{dcolumn}
\usepackage{subfigure}
\usepackage{comment}
\usepackage{multirow}
\usepackage{color}
\usepackage{bm}
\usepackage{mathrsfs}
\usepackage{amssymb}
\usepackage{amsfonts}
\usepackage{amsmath}
\usepackage{xspace}
\usepackage{float}
\usepackage{epstopdf}
\usepackage{tabularx}
\usepackage{longtable}
\usepackage[colorlinks=true, letterpaper=true, pdfstartview=FitV, linkcolor=blue, citecolor=blue, urlcolor=blue]{hyperref}
\usepackage[normalem]{ulem}
\usepackage[version=4]{mhchem}
\begin{document}
\title{Tentaclelike spectra and bound states in Hatano-Nelson chain with long-range impurity coupling}
\author{Xiaoming Zhao}
\email{xmzhao@ustb.edu.cn}
\affiliation{Department of Physics, Institute of Theoretical Physics and Shunde Innovation School, University of Science and Technology Beijing, Beijing 100083, China}
\affiliation{Key Laboratory of Multiscale Spin Physics (Ministry of Education), Beijing Normal University, Beijing 100875, China}
\author{Haiping Hu}
\email{hhu@iphy.ac.cn}
\affiliation{Beijing National Laboratory for Condensed Matter Physics, Institute of Physics, Chinese Academy of Sciences, Beijing 100190, China}
\affiliation{School of Physical Sciences, University of Chinese Academy of Sciences, Beijing 100049, China}
\begin{abstract}
In non-Hermitian systems, the energy spectra and eigenstates exhibit high sensitivity to boundary conditions, lattice geometries, and local impurities. In this paper, we study the effect of long-range impurity coupling, located far from the boundaries, on the paradigmatic non-Hermitian Hatano-Nelson model. Through exact analytical treatment, we reveal the intriguing tentacle-like spectral structures that emerge from the otherwise Bloch or non-Bloch spectra under periodic or open boundary conditions, respectively. We show that these spectral tentacles are associated with emergent bound states near the impurity, with their number determined by the coupling range. We further determine the localization length of these tentacled states using the transfer matrix. Our work indicates that the long-range impurity coupling cannot be treated as a mere perturbative effect and holds promise for state manipulations in non-Hermitian systems.
\end{abstract}
\maketitle

\section{Introduction}
Non-Hermitian systems exhibit a plethora of intriguing properties that go beyond those found in Hermitian systems \cite{coll1,coll2,coll4,coll5,coll6,coll7}. A notable example is the non-Hermitian skin effect (NHSE) \cite{NHSE1,NHSE2,NHSE3,NHSE4,NHSE5,NHSE6,NHSE7,NHSE8,NHSE9,NHSE10,NHSE11,NHSE12}, where an extensive number of eigenstates are pushed to the boundaries, which has been extensively studied experimentally\cite{ExpNHSE1,ExpNHSE2,ExpNHSE3,ExpNHSE4,ExpNHSE5,ExpNHSE6,ExpNHSE7,ExpNHSE8,ExpNHSE9}. In non-Hermitian systems, the energy spectra are highly dependent on the boundary conditions. Typically, under periodic boundary conditions (PBC), the spectra form closed loops on the complex plane \cite{hu_knot}, while under open boundary conditions (OBC), they form open arcs \cite{hu_graph,leech_graph}. This spectral sensitivity leads to promising applications such as loss-induced transparency \cite{LosIndTra}, unidirectional invisibility \cite{Uni12011,Uni22014} and high-performance senser \cite{senser,TopSen2020}.

\begin{figure}[ptb]
\centering
\includegraphics[clip,width=0.49\textwidth]{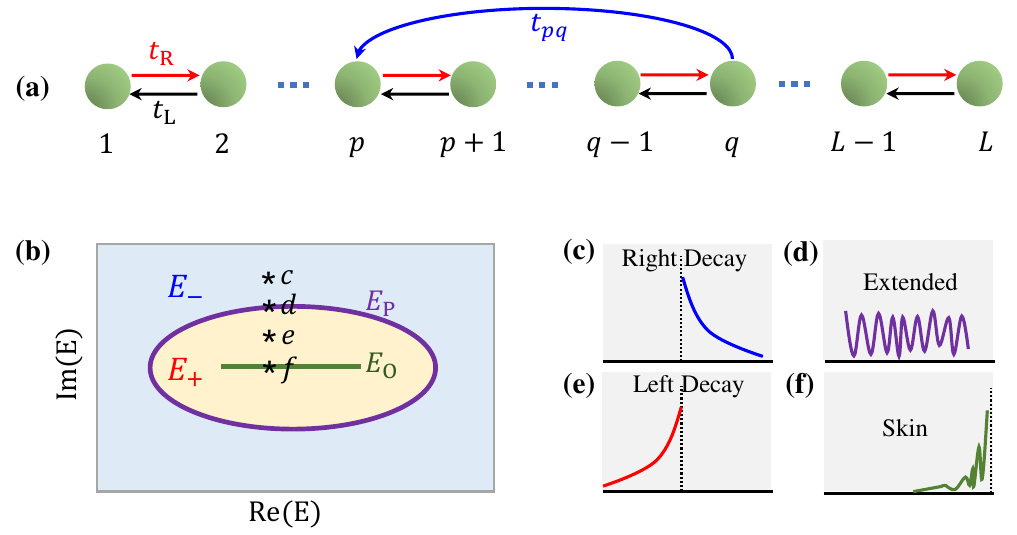}
\caption{Schematics of the tentacle-like bound states induced by long-range impurity coupling in non-Hermitian systems. (a) The Hatano-Nelson chain with a single long-range coupling from site $q$ to $p$ with $p>q$ or $p<q$. (b) The tentacle-like states that emerge from the continuum energy spectra under open (green) and periodic (purple) boundary conditions with varying the coupling strength $t_{pq}$. (c-f) Spatial profiles corresponding to the four distinct types of eigenstates, each indicated by star points labeled c, d, e and f in panel (b), respectively. Based on their locations on the complex plane, they correspond to right-decay bound state, extended Bloch state, left-decay bound states, and skin mode, respectively.}
\label{fig1}
\end{figure}

The distinct spectral structures associated with different boundary conditions are governed by non-Bloch band theory \cite{NHSE2,NHSE3}. Beyond boundary conditions, non-Hermitian systems with domain walls \cite{nh_domain1,nh_domain2,nh_domain3}, disorders \cite{nh_disorder1,nh_disorder2}, impurities \cite{nh_impurity1,nh_impurity2,nh_impurity3}, and varying lattice geometries \cite{hu_uniform,hu_nbloch,fangchen_gdse,zhangkai} have also garnered significant attention, displaying characteristics markedly different from Hermitian systems and altering our understanding of Bloch band theory. For impurity problems, there are two scenarios: local (e.g., onsite potential variations) and non-local impurities. Previous studies have shown that a local impurity can induce scale-free localization \cite{nh_impurity1,nh_impurity2,nh_impurity3,llh_sfl1,llh_sfl2,Gcx_sfl3}, where eigenstates accumulate near the impurity and the localization length depends linearly on the system size. However, the effect of non-local impurities (e.g., long-range impurity coupling as depicted in Fig. \ref{fig1}(a) on non-Hermitian systems remains unclear. Such long-range coupling can be implemented e.g., via tuning the spectrum or optical gain of pump light in photonic lattice or via flexible RLC-component wiring in electric circuits\cite{ExpNHSE1,ExpNHSE2,ExpNHSE4,Guo_NC,LRC_1,LRC_2,LRC_4}. From a perturbative perspective, one might intuitively guess that long-range impurity coupling far from the boundary would rarely impact the spectra and eigenstates because the skin modes reside at the boundaries. However, the perturbative intuition often does not hold true for non-Hermitian systems. The question remains: would such non-local impurities significantly reshape the energy spectra and eigenstates?

In this paper, we study the paradigmatic Hatano-Nelson chain with long-range impurity coupling. Through an exact analysis of the energy spectra and eigenstates, we reveal the emergence of spectral tentacles on the otherwise Bloch or non-Bloch spectra under PBC or OBC, respectively. We show that these tentacled states correspond to bound states near the impurity, with their number determined by the impurity coupling range and coupling strength, but not the system size. Furthermore, the localization properties of these tentacled states are determined using the transfer matrix method. Our results thus challenge the perturbative intuition, highlighting the significant differences between non-Hermitian and Hermitian systems, as well as between non-local and local impurities.

The rest of the paper is organized as follows. In Sec.~\ref{Hamil}, we review the band structure of the Hatano-Nelson model under different boundary conditions and present the analytical solutions of the energy spectra with long-range impurity coupling. In Sec.~\ref{SolSelfCon}, we derive the self-consistent equation that determines the eigenspectra and show the emergence of spectral tentacles under different boundary conditions. In Sec.~\ref{scaling}, we focus on the tentacled eigenstates induced by the long-range impurity coupling and discuss their different localization properties in distinct spectral regions on the complex plane. A summary is provided in Sec.~\ref{conclusion}.

\section{The Hatano-Nelson model with long-range impurity}\label{Hamil}

We consider the Hatano-Nelson model \cite{hnmodel} which exhibits the NHSE under OBC. An additional long-range impurity coupling is imposed on the $p, q$ sites, as depicted in Fig. \ref{fig1}(a). The Hamiltonian is expressed as
\begin{eqnarray}
H\!=\!H_{0}+t_{pq}c^{\dagger}_{p}c_{q}.
\label{Hami}
\end{eqnarray} 
Here, the second term represents the impurity coupling with strength $t_{pq}$. The first term is the Hatano-Nelson Hamiltonian
\begin{eqnarray}
H_{0}\!=\!\!\sum_{i=1}^{L-1}\!(t_{r}c^{\dagger}_{i+1}c_{i}\!+\!t_{l}c^{\dagger}_{i}c_{i+1}) 
\!+\!\delta(t_{r}c^{\dagger}_{1}c_{L}\!+\!t_{l}c^{\dagger}_{L}c_{1}),
\end{eqnarray}
where $c^{\dagger}_{i}$ ($c_{i}$) is the particle creation (annihilation) operator at the $i$-th site, $t_{l}$ ($t_{r}$) denotes the amplitude of the nearest-neighbor hopping to the left (right). For convenience, we set $t_{r}=1$ as the energy unit and $t_{l}=\alpha^{2}\in (0,1)$. $L$ is the chain length. The boundary condition is specified by $\delta=0$ (OBC) and $\delta=1$ (PBC), respectively. 

In the basis of $(c_{1},c_{2},...,c_{L})$, the Hamiltonian is represented by an $L\times L$ matrix. The energy spectra of $H$, denoted as $\sigma(H)$, are formally defined as the set of all complex $\lambda \in \textbf{C}$ such that $H-\lambda I$ is non-invertible. We are interested in the spectra when the impurity is introduced. To this end, we first review the spectral theory in the absence of the impurity $t_{pq}=0$. Due to the NHSE, the energy spectra exhibit distinct characteristics under different boundary conditions. For PBC, the energy spectra (i.e., the Bloch spectra) in the thermodynamic limit can be expressed as:
\begin{eqnarray}\label{EPBC}
\sigma(H_0)= e^{ik}+\alpha^{2}e^{-ik},
\end{eqnarray}
with $k\in[0,2\pi)$. The PBC spectra form an ellipse in the complex plane, parameterized by: $\left\{x^{2}/(1+\alpha^{2})^{2}+y^{2}/(1-\alpha^{2})^{2} =1\right\}$. In sharp contrast, under OBC, the energy spectra (i.e., the non-Bloch spectra) in the thermodynamic limit are given by 
\begin{eqnarray}\label{EOBC}
\sigma(H_0)=2\alpha\cos k,
\end{eqnarray}
which forms a line segment connectting the foci of the PBC ellipse. 

\begin{table}[!t]
\centering
\setlength{\tabcolsep}{2.4mm} 
\renewcommand{\arraystretch}{2.3}
\begin{tabular}{|c|c|c|c|c|}
  \hline
   \multicolumn{2}{|c|}{} & $p>q$ & $p=q$ & $p<q$ \\
  \hline
  \multirow{2}*{PBC} & $E_{+}$ &$\frac{\lambda_{1}^{p-q}-\lambda_{2}^{p-q}}{\lambda_{1}-\lambda_{2}}$ & $\varnothing$ & $\varnothing$ \\
  \cline{2-5}
     &$E_{-}$&$\frac{\lambda_{1}^{p-q}}{\lambda_{1}-\lambda_{2}}$&$\frac{1}{\lambda_{1}-\lambda_{2}}$&$\frac{\lambda_{2}^{p-q}}{\lambda_{1}-\lambda_{2}}$\\
  \hline
  \multirow{2}*{OBC} & $E_{+}$ & \multirow{2}*{$\frac{\lambda_{1}^{p-q}-\lambda_{1}^{p}/\lambda_{2}^{q}}{\lambda_{1}-\lambda_{2}}$} & \multirow{2}*{$\frac{1-\lambda_{1}^{p}/\lambda_{2}^{p}}{\lambda_{1}-\lambda_{2}}$} & \multirow{2}*{$\frac{\lambda_{2}^{p-q}-\lambda_{1}^{p}/\lambda_{2}^{q}}{\lambda_{1}-\lambda_{2}}$} \\
   \cline{2-2}
    & $E_{-}$ &   &   &   \\
  \hline
\end{tabular}
\caption{The expression of $d_{pq}(\lambda_1,\lambda_2)$ for different cases of $p$, $q$, and boundary conditions. The tentacle-like spectra which are determined by the self-consistent equation $d_{qp}=-1/t_{pq}$, are given by $E_{imp}=\lambda_1+\lambda_2$ with $\lambda_1\lambda_2=\alpha^2$. $E_{-}$/$E_{+}$ denotes the spectral region outside/inside of the ellipse of the Bloch spectra. The symbol $\varnothing$ denotes the absence of any solution.}
\label{Table-1}
\end{table}
Next, we turn to the case where the impurity coupling $t_{pq}\neq0$. Our question is whether the impurity brings significant changes to the energy spectra and eigenstates. The impurity is far from the boundaries, so we exam the scenario of finite $p,q$ in the thermodynamical limit $L\rightarrow\infty$. One of the main results of this work is that the energy spectra in the presence of the impurity contain two parts:
\begin{eqnarray}\label{SpH}
\lim_{L\rightarrow\infty}\sigma(H)=\sigma(H_{0})\cup \{\lambda\in \textbf{C}\backslash \sigma(H_{0}):d_{qp}=-\frac{1}{t_{pq}}\}.
\end{eqnarray}
Here $d_{qp}$ for different cases are listed in Table~\ref{Table-1}. $d_{qp}$ is defined as the $(p,q)$ entry of the resolvent or Green’s function. For PBC, $d_{p,q}=(L-\lambda I)_{p,q}^{-1}$ and for OBC, $d_{p,q}=(T-\lambda I)_{p,q}^{-1}$. Here $L$ and $T$ are Laurent matrix and Teoplitz matrix, respectively [See their definitions in Eq.~(A1) of the Appendix A], which account for the different boundary conditions. In the lattice-site basis, $t_{p,q}$ gives the propagator from site p to site q. In Eq.~(5), the first part comes from the Bloch or non-Bloch spectra of $H_0$, for either PBC or OBC. The second part is induced by the impurity and determined by the self-consistent equation $d_{qp}(\lambda)=-1/t_{pq}$, and the eigenenegy is given by $E_{imp}\equiv\lambda=\lambda_1+\lambda_{2}$. As will be shown later, it corresponds to tentacle-like bound states emerging from the Bloch or non-Bloch spectra.

\section{Tentacle-like spectra}\label{SolSelfCon}
\subsection{Self consistent equation under PBC and OBC}
Generally, The energy of the bound states $\lambda$ located on a unique ellipse in the complex plane characterized by
\begin{equation} \label{ellipse}
\lambda=\alpha^{2}\varrho ^{-1} e^{-i\theta}+\varrho e^{i\theta},
\end{equation}
with $\varrho>0$, $\theta\in[0,2\pi)$, and we set $\lambda_{1}=\alpha^{2}\varrho ^{-1} e^{-i\theta}$, $\lambda_{2}=\varrho e^{i\theta}$. It can be categorized into four intervals and defined as: 
\begin{equation}\label{InOut}
\begin{aligned}
 &\lambda\in E_{\text{P}}:\varrho=\alpha^2 \text{ or }\varrho=1,\\
  &\lambda\in E_{\text{O}}: \varrho=\alpha,\\
  &\lambda\in E_{+}:\varrho\in(\alpha^2,\alpha)\cup(\alpha,1),\\
  &\lambda\in E_{-}:\varrho\in(0,\alpha^2)\cup(1,+\infty),
\end{aligned}
\end{equation}
where the set $E_{\text{P}}$/$E_{\text{O}}$ represents the ellipse/line-segment defined in Eq.~(\ref{EPBC})/Eq.~(\ref{EOBC}), $E_{-}$ denotes the points situated outside of the ellipse $E_{\text{P}}$, $E_{+}$ denotes the points lie within $E_{\text{P}}$ but are not part of $E_{\text{O}}$. Based on the category of $\lambda$, we delve into the specifics of $d_{qp}(\lambda)$, which denotes the $(q,p)$ entry of the inverse matrices $(L-\lambda I)^{-1}$ under PBC and $(T-\lambda I)^{-1}$ under OBC (see Appendix \ref{appendixA}).

\begin{figure}[tbp]
\centering
\includegraphics[clip,width=0.49\textwidth]{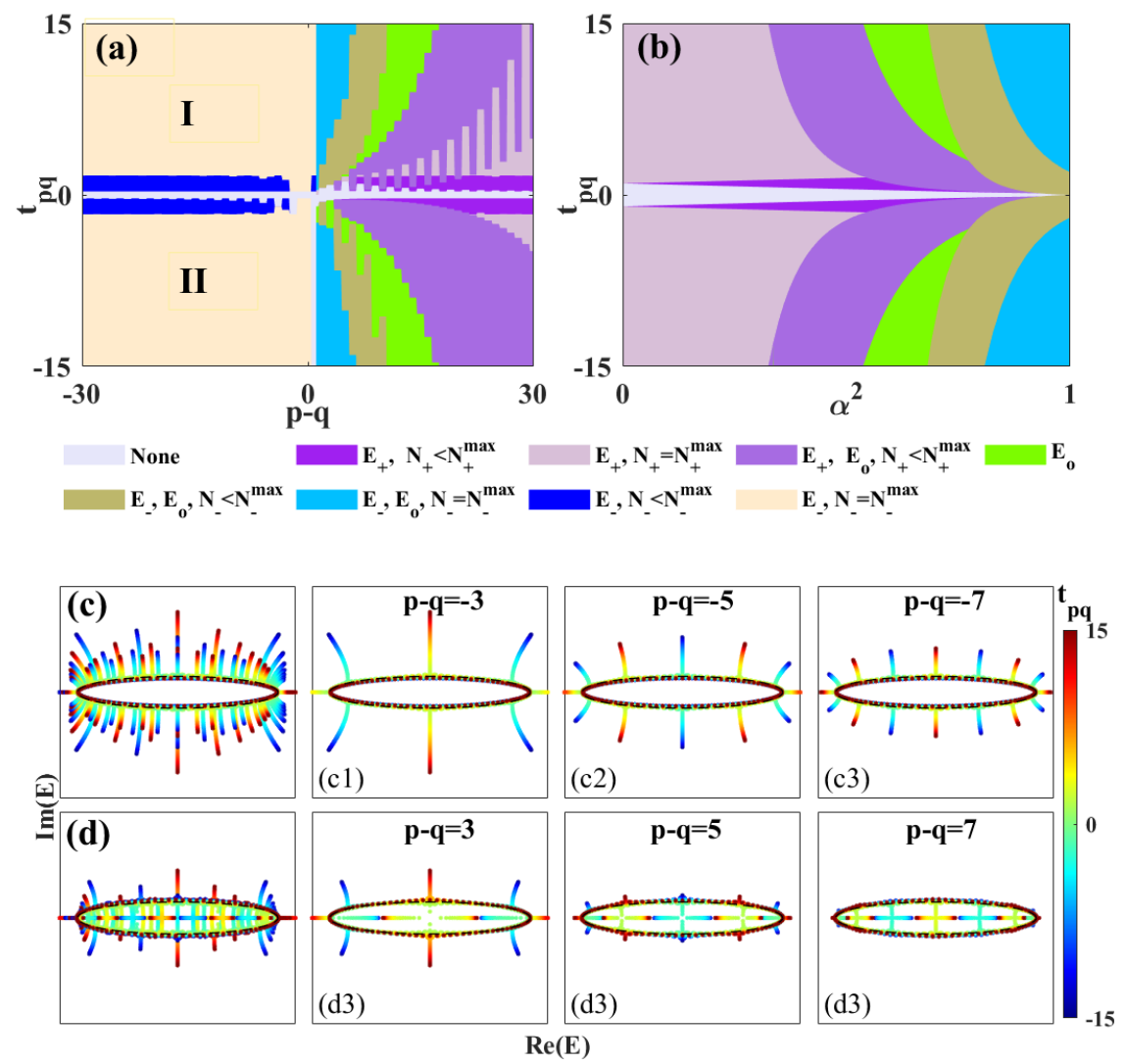}
\caption{The distribution of impurity energy presented in parameter space of the system under PBC. (a) $\alpha^{2}=$0.7. (b) $p-q=10$. (c-d) Energy spectral-flows perturbed by the single impurity in large size under PBC, which contain a ellipse and a few wings.  Here, we set $\alpha^{2}=0.7$, $L=1000$, the range of x/y-axes is $[-2,2]$. $p-q$ range from 3 to 10 in (c) and -10 to -3 in (d).}
\label{alphaPqXi}
\end{figure}

For example, under PBC, when $\mathbf{\lambda \in E_{-}}$, we have $\varrho\in(1, +\infty)$, which leads to $|\lambda_{1}| < 1$, $|\lambda_{2}| > 1$ and the symbol of $(L-\lambda I)^{-1}$ is
\begin{eqnarray}\label{PBCEMinus}
&&[a(t)-\lambda]^{-1}\nonumber\\
&&=-\frac{1}{\lambda_{2}}\frac{1}{\left( 1-\lambda_{1}/t\right)}\frac{1}{\left(1-t/\lambda_{2}\right)}\nonumber \\
&&=-\frac{1}{\lambda_{2}}\left(1+\frac{\lambda_{1}}{t}+\frac{\lambda_{1}^{2}}{t^{2}}+\cdots\right)\left(1+\frac{t}{\lambda_{2}}+\frac{t^{2}}{\lambda_{2}^{2}}+\cdots\right)\nonumber \\
&&=\frac{1}{\lambda_{1}-\lambda_{2}}\left[1+\sum_{n=1}^{\infty}\left( \frac{\lambda_{1}^{n}}{t^{n}}+\frac{t^{n}}{\lambda_{2}^{n}}\right)\right],
\end{eqnarray}
so the $(q-p)$-th Fourier coefficient of the symbol is written as
\begin{eqnarray}\label{DPBC1}
d_{qp}
&=\begin{cases}
\frac{\lambda_{1}^{p-q}}
{\lambda_{1}-\lambda_{2}}, \text{ } p> q\\\\
\frac{\lambda_{2}^{p-q}}
{\lambda_{1}-\lambda_{2}}, \text{ } p\leq q\\
\end{cases} 
\end{eqnarray}

Based on the same principle, we can get the functions $d_{qp}(\lambda)$ under PBC/OBC with $\mathbf{\lambda \in E_{+}}$ and $\mathbf{\lambda \in E_{-}}$, which are summarized in Table-\ref{Table-1}. The correction of the analytical outcomes under PBC and OBC is evidenced by their precise correspondence with the Hamiltonian diagonalization depicted in Fig.~\ref{EnergLocLength}(a) and (d), where the impurity energy obtained from the self-consistent equation (solid dot) matches well with the results of numerical calculations (hollow circle). Direct Hamiltonian diagonalization in large non-Hermitian systems often leads to unavoidable computational errors and is highly time-consuming, but our analytical self-consistent equations provide an efficient solution for single-impurity problem. 

\begin{figure}[tbp]
\centering
\includegraphics[clip,width=0.49\textwidth]{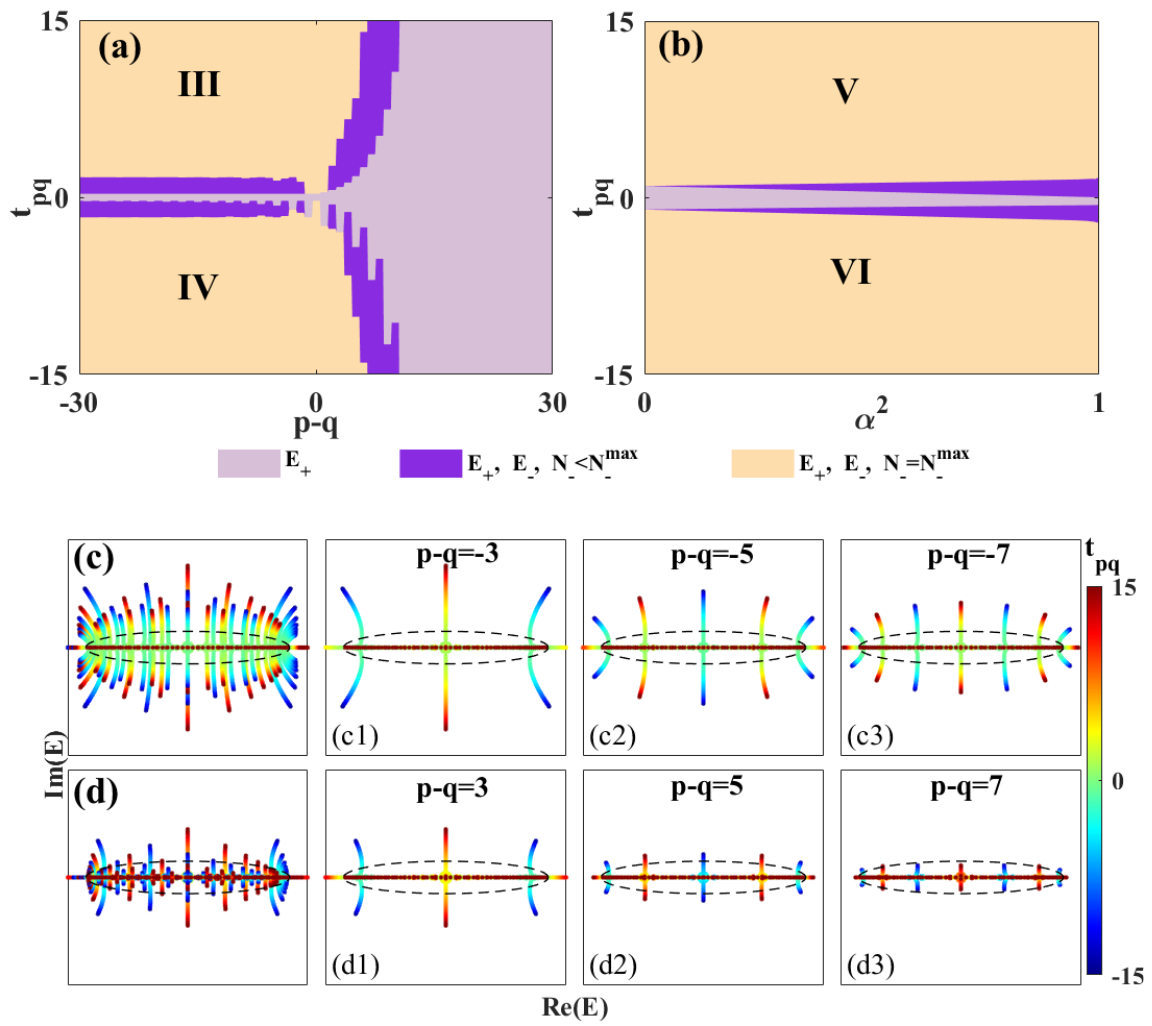}
\caption{The distribution of impurity energy $E_imp$ presented in parameter space under OBC for $\alpha^2=0.7$ in (a) and  p-q=-10 in (b).  (c), (d) Energy spectral-flows perturbed by the single impurity in large size under OBC, which contains a segment and a few wings. The shape depends on the value of p and q but not the p-q, we set p=60 here, and L $\alpha$, p-q vary as the same in Fig. 2(c), (d).}
\label{ImpurityPBC}
\end{figure}

\subsection{Energy distribution in parameter space}

Now, let's delineate the conditions satisfied by the boundaries between $E_{+}$ and $E_{-}$ in the parameter space spanned by $t_{pq}$, $p-q$ and $\alpha$. Referring to Eq.~(\ref{InOut}), $\varrho=1$ indicate the corresponding energy located on the ellipse $E_{\text{P}}$, leads to 
\begin{eqnarray}\label{Boundary}
\lambda_{1}=\alpha^{2}e^{-i\theta}, \text{  } \lambda_{2}=e^{i\theta},
\end{eqnarray}
which serves as the benchmark for determining the phase boundary. In details, for a given $p$, $q$, $\alpha$ and $t_{pq}$, the existence of $n$ solutions $\theta_{n}$ to self-consistent equation $d_{qp}(\lambda)=-1/t_{pq}$ implies that there are $n$ points located on the boundary $E_{P}$, otherwise, it is distributed in region $E_{+}$ or $E_{-}$ .

For example, we investigate the distribution of $E_{+}$ under PBC with $p>q$, where the number of solution depends on $p-q$ but not $p$, $q$ themselves. For $p-q=1$, $d_{pq}(\lambda)\!=\!1$, there is no solution except for $t_{pq}\!=\!-1$, this implies that $\sigma(H)$ contains no points in $E_{+}$, i.e., the long-range hopping $t_{pq}$ cannot add components to the spectrum in the interior of the ellipse.  For $p-q\!=\!2$, $d_{pq}(\lambda)\!=\!\lambda_{1}+\lambda_{2}\in E_{+}$ must satisfy $|t_{pq}|>1/(1+\alpha^{2})$, which means $\sigma(H)$ is not affected by sufficiently small impurities. 

The variation range of $N_{E_{-}}$ and $N_{E_{+}}$ in $t_{pq}$, $(p-q)$, $\alpha^{2}$ space under PBC are shown in Fig.~\ref{alphaPqXi}~(a) and (b). The energy spectral-flows perturbed by the single impurity in large size under PBC are depicted in Fig.~\ref{alphaPqXi}~(c) and (d). And the results for OBC are shown in Fig.~\ref{ImpurityPBC}~(a-d). In thermodynamic limit, with the increase of impurity strength $|t_{pq}|$, the energy spectral-flows contain a ellipse and a few wings under PBC, and it contain a segment and a few wings under OBC. This results are precisely what Eq.~(\ref{SpH}) describes. The results indicate that the long-range impurity coupling cannot be treated as a mere perturbative effect when non-Hermicity is introduced. This breaks the intuition in Hermitian systems, where a single impurity or domain wall typically induces only a few bound states, with the bulk energy spectrum remaining largely unaffected and the localization properties of eigenstates staying intact. For example, in topological phases of matter, an open boundary or a domain wall generally gives rise only to in-gap modes, as predicted by bulk topological invariants.

\section{Bound states induced by long-range impurity}\label{scaling}

\subsection{Localization in thermodynamic limit}
Based on the Hamiltonian in Eq.~(\ref{Hami}) and the eigen equation $H|\Psi\rangle=E|\Psi\rangle$ with $|\Psi\rangle=(\psi_{1},\psi_{2},\cdots,\psi_{L})^{T}$, the following non-local relations are obtained
\begin{equation}\label{NonLoc}
t_{R}\psi _{n-1}-E\psi _{n}+t_{L}\psi _{n+1}+\delta _{n,p}t_{pq}\psi _{q}=0
\end{equation}%
where $\delta _{n,p}=1$ ($\delta _{n,p}=0$) for $n=p$ ($n\neq p$) and because of spatial translational property in bulk $n\neq p$, we set the ansatz of wave function
\begin{eqnarray*}\label{ansatz}
|\Phi\rangle=\left(z^{p}, z^{p-1},\cdots, z, \eta z^{L},\eta z^{L-1},\cdots,\eta z^{p+1}\right) ^{T}
\end{eqnarray*}
Here, $z$ is complex, $\eta>0$ is real, $\eta=1$ for PBC and $\eta\neq1$ for OBC, and the bulk energy of the system read as
\begin{eqnarray}\label{Ez}
E=t_{R}z+t_{L}z^{-1}.
 \end{eqnarray}
 

Compared with the previous settings in Eq.~(\ref{ellipse}), we have $z=\varrho e^{i\theta}$ the Eq.~(\ref{NonLoc}) in the domain wall $n=p$, $n=p+1$ becomes
\begin{eqnarray*}\label{DomWal}
&&h_{p}(z)\equiv t_{L}+t_{L}\eta z^{L}+t_{pq}z_{pq}=0,\\
&&h_{p+1}(z)\equiv z^{L}-\eta z=0,
\end{eqnarray*}
where $z_{pq}=z^{L+q-p}$ for $p\geq q$, $z_{pq}=\eta z^{q-p}$ for $p<q$. For a certain $E$ in Eq.~(\ref{Ez}), there are two solutions $z_{1}=\varrho_{1} e^{i\theta}$, $z_{2}=\varrho_{2} e^{-i\theta}$ with $z_{1}z_{2}=\alpha^{2}$, and we can typically assume $|z_{1}|<|z_{2}|$. So the general solution of $H$ can be written as
\begin{eqnarray*}
|\Psi\rangle=\zeta_{1}|\Phi(z_{1})\rangle+\zeta_{2}|\Phi(z_{2})\rangle
 \end{eqnarray*}
and $\zeta_{1}$, $\zeta_{2}$ are complex value satisfied
$H_{p}\left[
\begin{array}{cc}
\zeta_{1} & \zeta_{2}%
\end{array}%
\right] ^{-1}=0$, 
where
\begin{equation}\label{Hp}
H_{p}=\left[
\begin{array}{cc}
h_{p}(z_{1}) & h_{p}(z_{2}) \\
h_{p+1}(z_{1}) & h_{p+1}(z_{2})%
\end{array}%
\right]
\end{equation}%
as a result, $\zeta_{1}$, $\zeta_{2}$, $z_{1}$, $z_{2}$, $\eta$ can be solved by $\det [H_{p}]=0$ and the normalizing condition $\langle\Psi|\Psi\rangle=1$  self consistently. While, in general, the self-consistent equation cann't be completely solved analytically.  While, we can discuss the localization property by the ratio form the second row of $H_{p}$ in Eq.~(\ref{Hp}) and decay coefficient $\xi$:
\begin{eqnarray}
\frac{\zeta_{1}}{\zeta_{2}}=\frac{z^{L}_{2}-\eta z_{2}}{z^{L}_{1}-\eta z_{1}},\text{ }\text{ }\text{ }\text{ }
\frac{1}{\xi}=\text{ln}\left|\frac{\psi_{n+1}}{\psi_{n}}\right|.
\end{eqnarray}

\begin{figure}[tb]
\centering
\includegraphics[clip,width=0.4\textwidth]{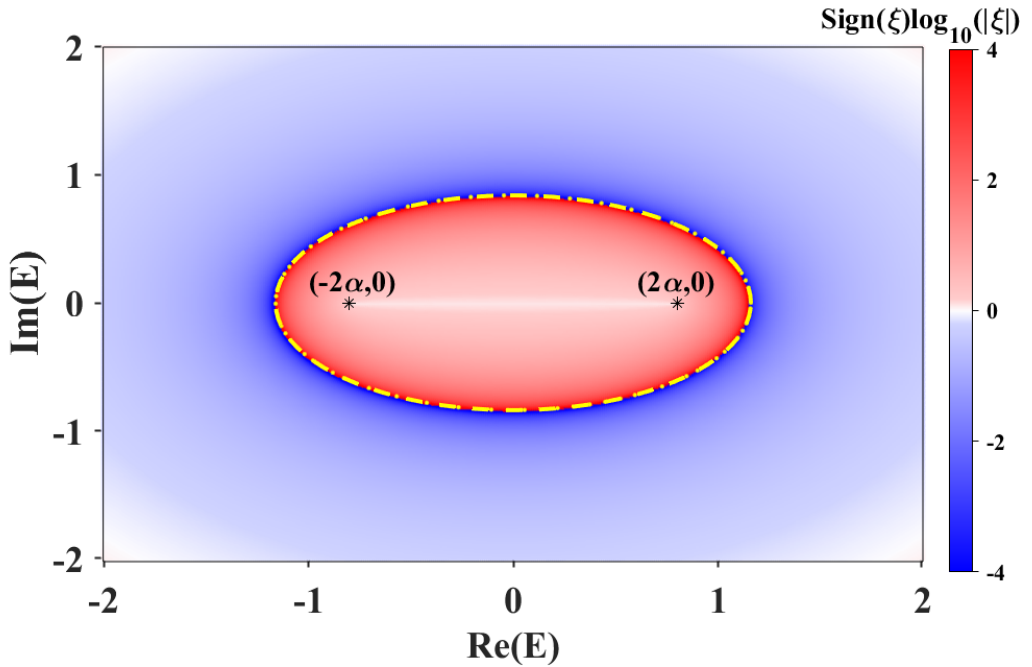}
\caption{Localization length calculated by the transfer matrix method for a arbitrarily energy $E$ in complex planet. $Sign(\xi)=+/-$, marked by red/blue series point, represents the corresponding wave function decay to left/right from the local center. The energy located on the segment between $(-2\alpha,0)$ and $(2\alpha,0)$ have the same localization length $-1/(2\text{ln}\alpha)$.  }
\label{localLength}
\end{figure}

In the context of the thermodynamic limit with $L\!\gg\!1$, we examines the localization characteristics at the boundary where $n=p$. This analysis encompasses the complete spectrum of energy values, in accordance with the definition provided by Eq.~(\ref{InOut}). 

i). For $E\in E_{\text{O}}$, $|z_{1}|=|z_{2}|=\alpha$, $\zeta_{1}/\zeta_{2}=1$ and $\xi\simeq1/\text{ln}\alpha$. This suggests that all states with energy levels confined within the range $[-2\alpha, 2\alpha]$ possess an identical finite localization length, corresponding to the NHSEs, as delineated in Fig.~\ref{fig1}(f). 

ii). For $E\in E_{+}$, $|z_{1}|\in(\alpha^2,\alpha)$, $|z_{2}|\in(\alpha,1)$, $\zeta_{1}/\zeta_{2}\simeq z_{2}/z_{1}$, $\xi\simeq-1/\text{ln}|\varrho_{2}|>0$, which implies the corresponding wave function decay to the left away from the local center, as depicted in Fig.~\ref{fig1}(e) and the designated red area in Fig.~\ref{localLength}. 

iii). For $E\in E_{\text{P}}$, $|z_{1}|=\alpha^{2}$, and $|z_{2}|=1$, $\zeta_{1}/\zeta_{2}\simeq (\eta-1)/\eta\alpha^{2}$, so we have $\xi\simeq -1/\text{ln}|\varrho_{2}|$ approaches to $+\infty$ under both OBC and PBC, these correspond precisely to the extended states, as depicted in Fig.~\ref{fig1}(d). 

iv). Similarly, for $E\in E_{-}$, we have $|z_{1}|=|\varrho_{1}|\in(0,\alpha^2)$, $|z_{2}|=\alpha^{2}/|z_{1}|\in(1,+\infty)$, so $|\zeta_{1}/\zeta_{2}|$ approaches to $+\infty$, $\xi\simeq -1/\text{ln}|z_{2}|=-1/\text{ln}|\varrho_{2}|<0$, the associated wave function is observed to decay to the right away from the local center, with this trend clearly illustrated in Fig.~\ref{fig1}(c) and the designated blue area in Fig.~\ref{localLength}. We conclude this discussion in the following table:
\begin{eqnarray}\label{comprehence}
\renewcommand\arraystretch{1.3}
\begin{tabular}{|c|c|c|c|}
  \hline
  & $|\zeta_{1}/\zeta_{2}|$ & $1/\xi$ &$|\Psi\rangle$\\
  \hline
$E\in E_{\text{O}}$ & 1& -2$\text{ln}\alpha$ & Localized \\
\hline
  $E\in E_{+}$ & $|z_{1}/z_{2}|$ & -$\text{ln}|\varrho_{2}|>0$ & Left-decay \\
  \hline
$E\in E_{\text{P}}$& $(\eta-1)/\eta\alpha^{2}$ & 0& Extended  \\
  \hline
 $E\in E_{-} $& +$\infty$ & -$\text{ln}|\varrho_{2}|<0$ & Right-decay \\
  \hline
\end{tabular}
\end{eqnarray}

\subsection{Scaling properties of the bound states} \label{TransMatMeth}
To study scaling properties of the bound states, we compute analytically the Lyapunov exponent using the transfer matrix method. The derivation of the transfer matrix of the model requires a more involved procedure. Based on the definition 
\begin{equation*}
\Phi _{n}=\left(
\begin{array}{c}
\psi _{n} \\
\psi _{n-1}%
\end{array}%
\right) ,A=\left(
\begin{array}{cc}
\frac{E}{\alpha^{2}} & -\frac{1}{\alpha^{2}} \\
1 & 0%
\end{array}%
\right),
\Gamma=\left(
\begin{array}{cc}
\frac{t_{pq}}{\alpha^ {2}} & 0 \\
0 & 0%
\end{array}%
\right),
\end{equation*}
the space evolution of the mode is written as
\begin{eqnarray}\label{GeneRelation}
\Phi _{n+1}
&=A\Phi _{n}+\delta _{n,p}\Gamma \Phi _{q}\equiv T_{n}\Phi _{n}
\end{eqnarray}%
where $T_{n}$ is the one-step real-space transfer matrix between $\Phi _{n+1}$ and $\Phi _{n}$. $T _{n}=A $ when $n\neq p$, and
\begin{equation}
T _{p}=
\begin{cases}
 \left(I-\Gamma A^{q-p-1}\right)^{-1}A, &p<q \\
\left(I+\Gamma A^{q-p-1}\right)A,&p\geq q
\end{cases}
\end{equation}%
when  $n=p$  (see Appendix \ref{AppendixTransMatrix}). Then the real-space transfer matrix of the complete system is
\begin{equation}\label{TransMetrComSy}
M=\prod\limits_{i=1}^n T_{i}=A^{L-p-1}T_{p}A^{p-1}
\end{equation}%
The localization length for a certain impurity state with large size systems is defined as the reciprocal of Lyapunov exponent $\xi(p,q,E_{\text{imp}})=\gamma^{-1}$ , $\gamma$ is the smallest positive eigenvalue of the matrix
\begin{equation}\label{LyaE}
\Gamma(p,q,\lambda)=\lim_{L\rightarrow\infty}\frac{1}{2L}\mathrm{ln}\left(M^{\dagger }M\right)
\end{equation}%

\begin{figure}[ptb]
\centering
\includegraphics[clip,width=0.45\textwidth]{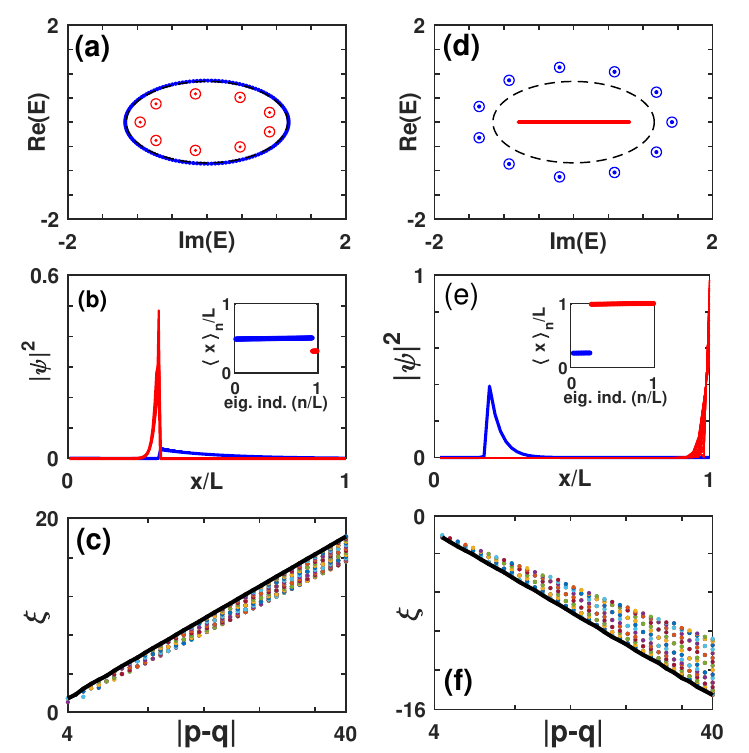}
\caption{Impurity states and its localization of the region $\textrm{I}$, $\textrm{II}$ in Fig.~\ref{alphaPqXi}(a) and $\textrm{III}$, $\textrm{IV}$ in Fig.~\ref{ImpurityPBC}(a). (a),(d) The energy spectrums for $|p-q|$=10 with $L=600$, $\alpha^{2}=0.2$, $t_{pq}=10$, $p=L/3$, red and blue circles denote the analytical results for the impurity states within and beyond the elliptical boundary $E_{\text{P}}$, respectively, while solid dots correspond to the relevant numerical data. (b),(e) The corresponding distribution of wave functions, the inset is the eigenvalue-resolved mean position. (c),(f) Show the localization length $\xi$ of all the $|p-q-1|$ impurity states, the scaling behavior is described by the liner black line defined by Max$|\xi|$. The positive (negative) localization length correspond to the right-decay states (left-decay states).} 
\label{EnergLocLength}
\end{figure}

From the Oseledets ergodic theorem \cite{Oseledets1968, Madabusi1979}. the above multiplication of transfer matrices is converged when $L\rightarrow\infty$. In principle, given the transfer matrix, one could directly compute the matrix product in Eq.~(\ref{LyaE}), and hence the Lyapunov exponents, as a function of N. However, in practice, such a numerical matrix multiplication and diagonalization is usually plagued by numerical rounding and overflow errors associated with the finite precision of the floating point representation of real numbers. In order to circumvent these issues, we perform a QR decomposition \cite{Ruelle1985,MacK1993} after every step involving a matrix multiplication. The QR decomposition of the first transfer matrix is $T_{1}=Q_{1}R_{1}$, where $Q_{1}$ is unitary and $R_{1}$ is upper triangular with real, positive diagonal entries, sorted in descending order. And with the definition $T'_{n+1}\equiv T_{n+1}Q_{n}=Q_{n+1}R_{n+1}$ with $Q_{n}^{\dagger}Q_{n}=1$, we get 
\begin{eqnarray}\label{QRDec}
\Gamma(p,q,\lambda)=\prod_{n=L}^{1}R_{n}^{\dagger}\prod_{m=1}^{L}R_{m}
\end{eqnarray}%
and the Lyapunov exponents 
\begin{eqnarray}\label{LyapunovExp}
\gamma=\text{Min}\{\gamma_{i}\}=\text{Min}\left\{\lim_{L\rightarrow\infty}\frac{1}{L}\sum_{n=1}^{L}\mathrm{ln}(R_{n,i})\right\}
\end{eqnarray}%
where $R_{n,i}$ is the $i$-th diagonal entries of $R_{n}$ with $i=1,2$.

As illustrative examples, we concentrate on several distinct regions within the parameter space for analytical scrutiny, i.e., $\textrm{I}$, $\textrm{II}$ in Fig.~\ref{alphaPqXi}(a) and $\textrm{III}$, $\textrm{IV}$ in Fig.~\ref{ImpurityPBC}(a), where $E_{imp} \in E_{-}$ with $\varrho\in(1, +\infty)$ and $|\lambda_{1}| < 1<|\lambda_{2}|$ can always be satisfied. When $p+q$ is large enough, the self-consistent equations in Eq.~(\ref{PBCEMinus}) and Eq.~(\ref{DOBCAp1}) with $p<q$ are both simplified to $\lambda_{2}^{p-q-1}\simeq1/t_{pq}$, the solutions of the equation are
\begin{eqnarray}\label{SimplifSCE}
(\widetilde{\varrho},\widetilde{\theta}_{n})\simeq \left(\sqrt[q-p+1]{|t_{pq}|}, \frac{2\pi}{q-p+1}n\right)
\end{eqnarray}
with $|t_{pq}|>1$ and $n=0,1,\cdots,q-p$. This means in thermodynamic limit $L\gg1$, whether the lattice chain is PBC or OBC, the hopping from site-$q$ to site-$p$ lead to $q-p+1$ impurity energies $\widetilde{E}_{n}=\frac{\alpha^{2}}{\widetilde{\varrho}e^{i\widetilde{\theta}_{n}}}+\widetilde{\varrho}e^{i\widetilde{\theta}_{n}}$,
which are all located on the same epsilon outside $E_{\text{PBC}}$. Form the result in last row of Eq.~(\ref{comprehence}), it is clear that all the impurity states have almost the same eigen wavefunction and, of course, the same localization length
\begin{eqnarray}\label{Line}
\xi=-1/\text{ln}|\widetilde{\varrho}|=-(q-p+1)/\text{ln}(t_{pq})
\end{eqnarray}
So, we have $\xi\varpropto q-p$, which implies the impurity induced eigenstates profile
respects a scale invariance. 

This feature can be demonstrated by a eigenvalue-resolved mean position 
\begin{eqnarray}\label{MeanPos}
\langle x\rangle_{n}=\sum^{L}_{m=1}\frac{m|\Psi_{n,m}|^{2}}{|\Psi_{n,m}|^{2}},
\end{eqnarray}
which does not change with the increase of $L$ when the ratio $p/L$ remains constant, as shown in Fig.~\ref{EnergLocLength}(b)(e). Additionally, we can see from Fig.~\ref{EnergLocLength}(c)(f) that the localization length increasing proportionally to the coupling distance $p-q$. This contrasts sharply from the scale-free localization observed in systems perturbed by a solitary onsite impurity where the localization length is proportional to the size of the entire system, and from the NHSEs states where the localization length remains invariant despite changes of system size. In general, each eigenstate possesses a localization length, with different localized states having distinct lengths. Consequently, a sector-like area is formed in Fig.~\ref{EnergLocLength}(c)(f), the maximum localization length to characterize the localization of the eigenstate. When $\alpha$ is sufficiently small, this sector-like area condenses into a single line. The findings indicate that the segmental scale-free bound states, induced by impurities, are ubiquitously present in the 1D NH chain.

\section{summary}\label{conclusion}
In this work, we have identified several crucial findings regarding the behavior of single long-range nonreciprocal coupling in 1D lattice systems. Initially, the single nonreciprocal coupling leads to the emergence of tentacle-like spectral as the coupling strength is gradually increased, the number of these wings precisely corresponds to the coupling distance $|p-q|$ within the lattice. Consequently, the energy spectra exhibit an elliptical/segmental shape with multiple wings under PBC/OBC. Simultaneously, the eigenstates exhibit a pronounced segmental localization, with the localization length increasing proportionally to the coupling distance. This contrasts sharply from the scale-free localization and the NHSEs states. Besides, the distribution of impurity energies in complex space express different decay behavior. Our findings are firmly established within a rigorous analytical framework, refined by the accuracy of the transform matrix method, providing a thorough grasp of the system’s intricate localization characteristics. Our work introduces an analytical method with distinct advantages for determining the energy spectra of non-Hermitian systems with impurities. It demonstrates that long-range impurity coupling is not merely a perturbation, and has promising implications for state manipulation in such systems. Experimentally, long-range non-reciprocal coupling can be simulated across various platforms, such as by tuning the spectrum or optical gain of pump light in photonic lattices, or through flexible RLC-component wiring in electrical circuits.

\begin{acknowledgments}
This work is supported by National Key Research and Development Program of China (Grant No. 2023YFA1406704 and 2022YFA1405800). X. Z. is supported by Fundamental Research Funds for the Central Universities, China (Grant No. FRF-TP-22-098A1), Guangdong Basic and Applied Basic Research Foundation (Grant No. 2023A1515110081) and Open Fund of Key Laboratory of Multiscale Spin Physics (Ministry of Education), Beijing Normal University (Grant No. SPIN2024K01). H. H. is also supported by the National Natural Science Foundation of China (Grant No. 12474496). 
\end{acknowledgments}

\appendix 
\section{The expression of $d_{qp}$ Under PBC/OBC}\label{appendixA}

Indeed, the Hamiltonian $H$ corresponds to finite Toeplitz matrices, a subject of study by Schmidt and Spitzer \cite{ToeAndLaur}. We demonstrate that the eigenvalue problem is effectively addressed with the assistance of Laurent matrices $L(a)$ and infinite Toeplitz matrice $T(a)$ under PBC and OBC, respectively, where
\begin{eqnarray}\label{ToeplitzM}
  L(a)=(a_{i-j})^{\infty}_{i,j=-\infty},\text{ }\text{ }\text{ }
 T(a)=(a_{i-j})^{\infty}_{i,j=1},
\end{eqnarray}
with $a_{i-j=1}=1$, $a_{i-j=-1}=\alpha^{2}$ and $a_{i-j\neq\pm1}=0$. We can express $H(a)$, $T(a)$ and $L(a)$ using the symbol 
 \begin{equation}
 a(t)=t+\alpha ^{2}t^{-1}, 
 \end{equation}
 with $t=e^{i\theta}$, $\theta\in[0,2\pi)$. As a result, the expression of $d_{qp}$, which represent the $(q,p)$ entry of $[L(a)-\lambda I]^{-1}$ or $[T(a)-\lambda I]^{-1}$, can also be described by the continuous function \cite{LargToe1,LargToe2}: 
\begin{equation}\label{factorization}
  \left[a(t)-\lambda\right]^{-1}=t(t-\lambda_{1})^{-1}(t-\lambda_{2})^{-1}.
\end{equation}
with $\lambda=\lambda_{1}+\lambda_{2},\alpha ^{2}=\lambda_{1}\lambda_{2}$.

As we have claimed, the parameter $\lambda$ exhibits a monotonic increase with respect to $\varrho $ for values in the range for $[\alpha,+\infty)$. Under this condition, it is always possible to establish that  $|\lambda_{1}|\leq|\lambda_{2}|$, $\lambda_{1}$ and $\lambda_{2}$ are defined as $\lambda_{1}=\alpha^{2}\varrho ^{-1} e^{-i\theta}$, $\lambda_{2}=\varrho e^{i\theta} $. Here, we delve into the specifics of
$d_{qp}$, which denotes the $(q,p)$ entry of the inverse matrices $(L-\lambda I)^{-1}$ under PBC and $(T-\lambda I)^{-1}$ under OBC. 

$\mathbf{\textcolor[rgb]{0.00,0.07,1.00}{Case 1:}}$ Under PBC, $\mathbf{\lambda \in E_{+}}$ with $\varrho\in(\alpha, 1)$. In this case, we have $|\lambda_{1}| < 1$ and $|\lambda_{2}| < 1$ and thus
\begin{eqnarray}\label{PBCEPlus1}
&&[a(t)-\lambda]^{-1}\nonumber\\
=&&\frac{1}{t}+\frac{1}{t^{2}}(\lambda_{1}+\lambda_{2})+\frac{1}{t^{3}}(\lambda_{1}^{2}+\lambda_{1}\lambda_{2}+\lambda_{2}^{2})+\cdots \nonumber \\
=&&\sum_{m=1}^{\infty}\frac{1}{t^{m}}(\sum_{n=0}^{m-1}\lambda_{1}^{n}\lambda_{2}^{m-1-n})
\end{eqnarray}
The $(q-p)$-th Fourier coefficient of the inverse function $\left[a(t)-\lambda\right]^{-1}$ is zero under $p\leq q$, while when $p>q$ it can be interpreted as
\begin{eqnarray} \label{PBCEPlusAppendix}
d_{qp}(\lambda)
&=&\sum_{s=1}^{p-q}\lambda_{1}^{s-1}\lambda_{2}^{p-q-s}\nonumber \\
&=&\frac{\lambda_{2}^{2(p-q)}-\alpha^{2(p-q)}}{\lambda_{2}^{p-q-1}(\lambda_{2}^{2}-\alpha^{2})}\nonumber \\
&=&\frac{(\lambda_{2}^{p-q}+\alpha^{p-q})(\lambda_{2}^{p-q}-\alpha^{p-q})}{\lambda_{2}^{p-q-1}(\lambda_{2}^{2}-\alpha^{2})}\nonumber \\
&=&\frac{\lambda_{2}^{p-q}-\lambda_{1}^{p-q}}{\lambda_{2}-\lambda_{1}}
\end{eqnarray}

$\mathbf{\textcolor[rgb]{0.00,0.07,1.00}{Case 2:}}$ PBC, $\mathbf{\lambda \in E_{-}}$ with $\varrho\in(1, +\infty)$.
Then we have $|\lambda_{1}| < 1$ and $|\lambda_{2}| > 1$ and thus we get
\begin{eqnarray}\label{PBCEMinus}
&&[a(t)-\lambda]^{-1}\nonumber\\
&&=-\frac{1}{\lambda_{2}}\frac{1}{\left( 1-\lambda_{1}/t\right)}\frac{1}{\left(1-t/\lambda_{2}\right)}\nonumber \\
&&=-\frac{1}{\lambda_{2}}\left(1+\frac{\lambda_{1}}{t}+\frac{\lambda_{1}^{2}}{t^{2}}+\cdots\right)\left(1+\frac{t}{\lambda_{2}}+\frac{t^{2}}{\lambda_{2}^{2}}+\cdots\right)\nonumber \\
&&=\frac{1}{\lambda_{1}-\lambda_{2}}\left[1+\sum_{n=1}^{\infty}\left( \frac{\lambda_{1}^{n}}{t^{n}}+\frac{t^{n}}{\lambda_{2}^{n}}\right)\right]
\end{eqnarray}
Following the same derivation process as in Eq.~(\ref{PBCEPlusAppendix}), the $(q-p)$-th Fourier coefficient of the inverse function $\left[a(t)-\lambda\right]^{-1}$ can be seen as
\begin{eqnarray}\label{DPBC1}
d_{qp}
&=\begin{cases}
\frac{\lambda_{1}^{p-q}}
{\lambda_{1}-\lambda_{2}}, \text{ } p> q\\\\
\frac{\lambda_{2}^{p-q}}
{\lambda_{1}-\lambda_{2}}, \text{ } p\leq q\\
\end{cases} 
\end{eqnarray}

$\mathbf{\textcolor[rgb]{0.00,0.07,1.00}{Case 3:}}$ Under OBC, for $\varrho\in(1, +\infty)$. Then we have $|z_{1}| < 1<|z_{2}|$, and all the energy points are located outside the line segment between the two foci of the ellipse $a(t)$, i.e., $E_{O}=[-2\alpha,2\alpha]$. The representation
\begin{eqnarray}\label{aPM1}
 &&a(t)-\lambda \nonumber \\
=&&t^{-1}(t^2-\lambda t+\alpha^2)\nonumber\\
=&&t^{-1}(1-\lambda_{1})(t-\lambda_{2})
\end{eqnarray}
is a Wiener–Hopf factorization, and from
 $T^{-1}(a-\lambda)=T(a_{+}^{-1})T(a_{-}^{-1})$ we therefore deduce that
\begin{eqnarray}\label{aPM1}
 a_{+}=t-\lambda_{2},\text{ }\text{ }a_{-}=1-\lambda_{1}/t
\end{eqnarray}
so taking into account that
\begin{eqnarray}\label{aPM2}
 &&a_{+}^{-1}=-\frac{1}{\lambda_{2}}\left(1+\frac{\lambda_{2}}{t}+\frac{\lambda_{2}^{2}}{t^{2}}+\cdots\right),\nonumber\\
 &&a_{-}^{-1}=1+\frac{t}{\lambda_{1}}+\frac{t^{2}}{\lambda_{1}^{2}}+\cdots
\end{eqnarray}
we arrive at representation
\begin{eqnarray}\label{TOBCAp}
& &T^{-1}(a-\lambda)
 =T(a_{+}^{-1})T(a_{-}^{-1})\nonumber\\
 &=&-\frac{1}{\lambda_{2}}
 \left(
\begin{array}{llll}
1 &  &   & \\
1/\lambda_{2} & 1 &   \\
 1/\lambda_{2}^{2} & 1/\lambda_{2} &  &  \\
 \cdots & \cdots & \cdots &  \\
\end{array}%
\right)
 \left(
\begin{array}{llll}
1 & \lambda_{1} & \lambda_{1}^{2}  &\cdots \\
 & 1 & \lambda_{1}& \cdots \\
   & & 1 & \cdots \\
 &  & &\cdots  \\
\end{array}%
\right),\nonumber
\end{eqnarray}
so the $(q,p)$ entry of $(T-\lambda I)^{-1}$ read as
\begin{eqnarray}\label{DOBCAp1}
d_{qp}
&=&-\sum_{s=1}^{\mathrm{min}\{p,q\}}\frac{\lambda_{1}^{p-s}}{\lambda_{2}^{q+1-s}}\nonumber\\
&=&
\begin{cases}
\frac{\lambda_{1}^{p-q}-\lambda_{1}^{p}/\lambda_{2}^{q}}{\lambda_{1}-\lambda_{2}}, \text{ } p\geq q \\\\
\frac{\lambda_{2}^{p-q}-\lambda_{1}^{p}/\lambda_{2}^{q}}{\lambda_{1}-\lambda_{2}}, \text{ } p<q 
\end{cases}.
\end{eqnarray}
In particular,
\begin{equation}
\begin{aligned}
&d_{11}=-\frac{1}{\varrho e^{i\theta}},
d_{12}=-\frac{\alpha^{2}}{\varrho^{2}e^{2i\theta}},\\
&d_{21}=\frac{1}{\varrho^{2}e^{2i\theta}},
d_{22}=-\frac{1}{\varrho e^{i\theta}}\left(\frac{\alpha^{2}}{\varrho^{2}e^{2i\theta}}+1\right).
\end{aligned}
\end{equation} 

In general, for $\varrho\in(\alpha, +\infty)$, the condition $|z_{1}|<|z_{2}|$ is always true, and the expressions for $d_{qp}(\lambda)$ found in Eq.~(\ref{DOBCAp1}) can be extended by analyticity to all $\lambda\in \textbf{C}\backslash E_{O}$.

\section{Transform matrix of the impurity systems}\label{AppendixTransMatrix}

In the bulk, the following non-local relations are obtained
\begin{equation}
t_{R}\psi _{n-1}-E\psi _{n}+t_{L}\psi _{n+1}+\delta _{n,p}\varepsilon\psi _{q}=0
\end{equation}%
$\varepsilon$ is the long-range hopping amplitute from q-th to p-th lattice and thus the transfer matrix of the system cannot be reduced to a product of A matrices. For these reasons, the derivation of the transfer matrix of the model requires a more involved procedure. We set
\begin{equation*}
\Phi _{n}=\left(
\begin{array}{c}
\psi _{n} \\
\psi _{n-1}%
\end{array}%
\right) ,A=\left(
\begin{array}{cc}
\frac{E}{\alpha^{2}} & -\frac{1}{\alpha^{2}} \\
1 & 0%
\end{array}%
\right),
\Gamma=\left(
\begin{array}{cc}
\frac{\varepsilon}{\alpha^{2}} & 0 \\
0 & 0%
\end{array}%
\right)
\end{equation*}
and the space evolution of the mode is controlled by the following equation:%
\begin{eqnarray}\label{GeneRelation}
\Phi _{n+1}&&=A\Phi _{n}+\delta _{n,p}\left(
\begin{array}{c}
\frac{\varepsilon}{\alpha^{2}}\psi _{q} \\
0%
\end{array}\right)\nonumber\\
&&=A\Phi _{n}+
\delta _{n,p}\left(
\begin{array}{cc}
\frac{\varepsilon}{\alpha^{2}} & 0 \\
0 & 0%
\end{array}%
\right)
\left(
\begin{array}{c}
\psi _{q} \\
\psi _{q-1}%
\end{array}%
\right)\nonumber\\
&&=A\Phi _{n}+\delta _{n,p}\Gamma \Phi _{q}
\end{eqnarray}%
so the relationship between $\Phi _{p+1}$ and $\Phi _{p}$ is unique compared to others, which can be discussed in the following two cases.
\bigskip

$\mathbf{\textcolor[rgb]{0.00,0.07,1.00}{Case\text{ }1:}\text{ }p<q}$.
Based on Eq.~(\ref{GeneRelation}), for $n=q,$
\begin{equation}\label{NEqP4}
\Phi _{q}=A^{q-p-1} \Phi _{p+1}
\end{equation}%
and for $n=p,$
\begin{eqnarray}\label{NEqP2}
\Phi _{p+1}=A\Phi _{p}+\Gamma A^{q-p-1} \Phi _{p+1}
\end{eqnarray}%
so we have
\begin{equation}
\Phi_{p+1}=\left(I-\Gamma A^{q-p-1}\right)^{-1}A\Phi _{p}
\end{equation}%

\bigskip
$\mathbf{\textcolor[rgb]{0.00,0.07,1.00}{Case \text{ }2:}\text{ }p\geq q.}$
For $n=q,$
\begin{equation}\label{NEqP4}
\Phi _{q}=A^{-(p-q)} \Phi _{p}
\end{equation}%
and for $n=p,$
\begin{eqnarray}\label{NEqP2}
\Phi _{p+1}=A\Phi _{p}+\Gamma A^{q-p} \Phi _{p}
\end{eqnarray}%
so we have
\begin{equation}
\Phi_{p+1}=\left(I+\Gamma A^{q-p-1}\right)A\Phi _{p}
\end{equation}%

\bigskip
With the definition
\begin{equation}
B=
\begin{cases}
 \left(I-\Gamma A^{q-p-1}\right)^{-1}A, &p<q \\
\left(I+\Gamma A^{q-p-1}\right)A,&p\geq q
\end{cases}
\end{equation}%
we have%
\begin{equation}
\Phi _{m}=
\begin{cases}
 A^{m-1}\Phi _{1},&m\leq p \\
A^{m-p-1}BA^{p-1}\Phi _{1},&m>p
\end{cases}
\end{equation}
Then the transfer matrix of the system is
\begin{equation}
T_{m}=
\begin{cases}
A^{m-1},&m\leq p \\
A^{m-p-1}BA^{p-1},&m>p
\end{cases}.
\end{equation}%
Specially, for $m=L+1$ we have $\Phi _{L+1}=T_{L}\Phi _{1}$, i.e.,
\begin{equation}
\left(
\begin{array}{c}
\psi _{L+1} \\
\psi _{L}%
\end{array}%
\right) =\left(
\begin{array}{cc}
T_{11} & T_{12} \\
T_{21} & T_{22}%
\end{array}%
\right) \left(
\begin{array}{c}
\psi _{1} \\
\psi _{0}%
\end{array}%
\right)
\end{equation}%
For OBC $\psi _{L+1}=\psi _{0}=0,$
which implies
\begin{equation}
T_{11}=0,\psi _{L}=T_{21}\psi _{1}.
\end{equation}
and for PBC $\psi _{L+1}=\psi _{1},\psi _{L}=\psi
_{0},$ which implies%
\begin{equation}
T^{2}=1.
\end{equation}%

\end{document}